\documentclass[twocolumn,preprintnumbers,amsmath,amssymb]{revtex4-1}
\usepackage{epsfig}
\usepackage{color}
\usepackage{amsmath}

\begin{document}
\title{Heterogeneous versus homogeneous crystal nucleation in hard spheres}

\author{Jorge R. Espinosa$^{1,4}$, Carlos Vega$^1$, Chantal Valeriani$^2$, Daan Frenkel$^3$ and Eduardo Sanz$^1$}
\affiliation{
$^1$ Departamento de Quimica Fisica,
Facultad de Ciencias Quimicas, Universidad Complutense de Madrid,
28040 Madrid, Spain\\
$^2$ Departamento de Estructura de la Materia, Fisica Termica y Electronica, Facultad de Ciencias Fisicas, Universidad Complutense de Madrid, 28040 Madrid, Spain\\
$^3$ Department of Chemistry, University of Cambridge, Lensfield Road, Cambridge CB2 1EW, United Kingdom\\
$^4$ Maxwell Centre, Cavendish Laboratory, Department of Physics, University of Cambridge, United Kingdom}

\date{\today}

\begin{abstract}

Hard-sphere model systems are well-suited in both experiment and simulations to investigate fundamental aspects of the crystallization of fluids.  
In experiments on colloidal models of hard-sphere fluids,  the fluid is unavoidably at contact with the walls of the sample cell, where heterogeneous 
crystallization may take place. In this work we use simulations to investigate the competition 
between homogeneous and heterogeneous crystallization. We report simulations of wall-induced nucleation for different confining walls. Combining the results of these simulations with earlier studies  of  homogeneous allows us to 
asses the competition between homogeneous and heterogeneous nucleation as a function of wall type, fluid density and the system size. 
On flat walls, heterogeneous nucleation will typically overwhelm homogeneous  nucleation. However, even for surfaces randomly coated with spheres with a diameter that was  some three times larger than that of the fluid spheres -- as  has been used in some experiments -- heterogeneous nucleation is likely to be dominant for volume fractions smaller than $\sim$ 0.535. 
Only for a disordered coating that has the same structure as the liquid holds promise did we find the nucleation was likely to occur in the bulk. Hence, such coatings might be used to suppress heterogeneous nucleation in experiments. 
Finally, we report the apparent homogeneous nucleation rate taking into account the formation of crystallites both in the 
bulk and at the walls. We find that the apparent overall nucleation rates coincides with those reported in ``homogeneous nucleation'' experiments. 
	This suggests that heterogeneous nucleation at the walls could partly explain the large discrepancies
	found between experimental measurements and simulation estimates of the homogeneous nucleation rate.

\end{abstract}
\maketitle

\section{Introduction}

Crystallization is ubiquitous, and a quantitative understanding of this phenomenon is important for all processes, both basic and applied, where crystallization plays a role. 

The crystallization of a disordered assembly 
of building blocks (atoms, molecules, colloids, proteins...) 
must be initiated by the nucleation of a crystalline embryo somewhere in the system~\cite{kelton}.
The critical nucleus may either emerge  in the homogeneous bulk, or heterogeneously, 
on a surface or on an impurity \cite{kelton}. The latter mechanism is often faster because it reduces the surface free energy and thereby the nucleation barrier 
of  the emerging solid. 

In this work we examine the competition between homogeneous and heterogeneous
nucleation for a system composed of hard spheres, which is archetypal for the  
study of the 
fluid-to-crystal transition \cite{alder:1208,woodjac,N_1986_320_340,Nature_2001_409_1020}.  
The hard-sphere model was originally generated in computer simulations \cite{alder:1208,woodjac}, but
it could, to a good approximation, be realized in experiments on colloidal suspensions \cite{N_1986_320_340}. However, in these experiments 
the sample is unavoidably contained in a cell and  therefore crystallization may occur either in the bulk or on the cell walls. 

Here, we use computer simulations~\cite{JCP_1992_96_04655,Nature_2001_409_1020,searJPCM2007,anwarAnge2011,reviewMichaelides2016} 
 to asses the competition between homogeneous and heterogeneous crystallization in hard sphere fluids. 

As simulation data on homogeneous hard-sphere nucleation are readily available, we focus on the 
heterogeneous nucleation taking place on the walls confining the hard sphere fluid. 
The competition between homogeneous and heterogeneous nucleation depends on 
system size as the rates of homogeneous 
and heterogeneously nucleation clusters are proportional to the 
volume and  the area of the cell,  respectively. As a consequence the dominant nucleation
mechanism will depend on the system size and the volume fraction occupied by the
spheres, $\phi$. 

To assess the homogeneous-vs-heterogeneous competition we
perform estimates of the time required for a crystal to appear via either mechanism.  
The homogeneous nucleation rate, $J_{hom}$, has been previously estimated with a range of 
different simulation techniques such as umbrella sampling
\cite{Nature_2001_409_1020,filion:244115}, forward-flux sampling
\cite{filion:244115}, lattice mold \cite{latticemold}, or seeding \cite{seedingvienes}.  Here, we use 
a reasonably good fit for $J_{hom}(\phi)$ provided by seeding simulations \cite{seedingvienes}. 
For the heterogeneous nucleation on a flat wall , $J_{het}$, the rate had been  computed previously  for a single state point,  using umbrella sampling~\cite{auer2003line}.
In the present paper we report calculations of $J_{het}$ as a function of $\phi$ using ``brute-force''
molecular dynamics simulations. Both flat and coated walls are considered.

We find that, for a given system size, there is a crossover from heterogeneous to homogeneous 
nucleation as density increases. The crossover is due to the fact that the homogeneous
nucleation rate increases more sharply with density. Moreover, our simulations predict that, for 
walls that are either flat or randomly coated with spheres about three times larger than the fluid ones, 
suppression of heterogeneous nucleation is not viable  for 
$\phi < $ 0.535, unless  astronomically large samples are considered.

This raises the question if any wall coating can be designed that would make heterogeneous nucleation less important than homogeneous nucleation.
We find that an immobilized wall coating replicating on the structure of the fluid  could 
efficiently prevent heterogeneous nucleation. 
Finally, by combining homogeneous and heterogeneous nucleation rates we compute an apparent homogeneous 
nucleation rate that takes into account crystallites nucleated both in the bulk and at 
the cell walls. We find that the apparent rate matches reported experimental data for $J_{hom}$ \cite{PRE_1997_55_3054,sinn,schatzel}.
Our findings suggests that careful corrections for heterogeneous nucleation are needed when comparing 
experimental  and numerical estimates  of the homogeneous nucleation rate \cite{Nature_2001_409_1020,filion:244115}. 

\section{Results}

\subsection{Heterogeneous nucleation rate on flat walls}

The heterogeneous nucleation rate, $J_{het}$, is defined as the number of clusters that nucleate per unit 
time and area:
\begin{equation}
	J_{het}=\frac{1}{A t_{het}}.
	\label{ref}
\end{equation}
where $A$ is the area of the walls, and $t_{het}$ is the average time required for a critical crystal nucleus 
to appear at the walls.  

To obtain $J_{het}$ we performed Molecular Dynamics (MD) simulations in a box with a (quasi) flat hard wall
perpendicular to $z$ and periodic boundary 
conditions in the other two directions. 
The size of the flat hard wall is $11.5$x$11.5$ $\sigma^2$, 
where $\sigma$ denotes the hard sphere diameter. 

We performed simulations with 2548 fluid particles. 
We fixed the pressure with a barostat along the $z$ direction. 
To perform MD simulations (with the GROMACS  package~\cite{hess08}) we use the
pseudo-hard sphere potential~\cite{pseudoHS}, where the discontinuous hard sphere interaction is replaced by the expression: 

 \begin{equation}
\left\{
\begin{array}{rc}
50 \left( \frac{50}{49} \right)^{49} \epsilon \left[ \left( \frac{\sigma}{r} \right)^{50} -\left( \frac{\sigma}{r} \right)^{49} \right] + \epsilon; &  r< \left(\frac{50}{49}\right) \sigma  \\
0; & r \ge \left( \frac{50}{49} \right)\sigma
\end{array}
\right .
\label{potential_50_49}
\end{equation}  
This model mimics the equilibrium~\cite{pseudoHS,sigmoide} and dynamic 
~\cite{pseudoHS,avalanchesPHS,brownianavalanches} properties of  hard spheres. Approximating hard spheres by very steep but continuous potentials,
allows  us to use a constant timestep MD code, such as the one
implemented in GROMACS. These codes have the advantage that they allow
for highly efficient parallel simulations, which is important for the
large system sizes needed to study nucleation. In contrast, event driven
hard-sphere MD requires active load-balancing. Whilst parallel,
event-driven MD codes have been proposed (see, e.g. \cite{miller2004event}) 
we opted for the GROMACS approach, because it is versatile and has been extensively tested. 

In practise, the flat wall is built by 
positioning spheres with the same size as the fluid ones in a single-layer square lattice of 0.7 $\sigma$ lattice spacing. 
This makes a virtually flat slab whose particles interact with the fluid ones via the pseudo-hard sphere potential. 
We have
checked with specific cases that by reducing the lattice spacing of the spheres that form the wall to 0.4 $\sigma$ we obtain the same results. This confirms
that the wall is effectively flat already with a 0.7 $\sigma$ spacing. The interaction between the fluid and the wall 
spheres is given by the pseudo-hard sphere potential defined above.
All other simulation details are the same as in our previous publications using the pseudo-hard sphere potential~\cite{sigmoide,seedingvienes}.

\begin{figure}[htb]
\begin{center}
	\includegraphics[clip,width=0.4\textwidth]{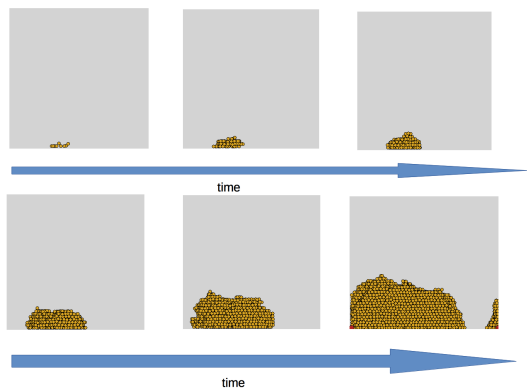}
	\caption{Sequence of snapshots showing a heterogeneous nucleation event at $\phi=0.52$. The box is filled with particles, but only 
	those detected to belong to the largest crystalline cluster according to a specific local-bond order 
	parameter~\cite{peterRoySoc2009,seedingvienes} are shown.}
\label{snap}
\end{center}
\end{figure}

\begin{figure}[htb]
\begin{center}
	\includegraphics[clip,width=0.4\textwidth]{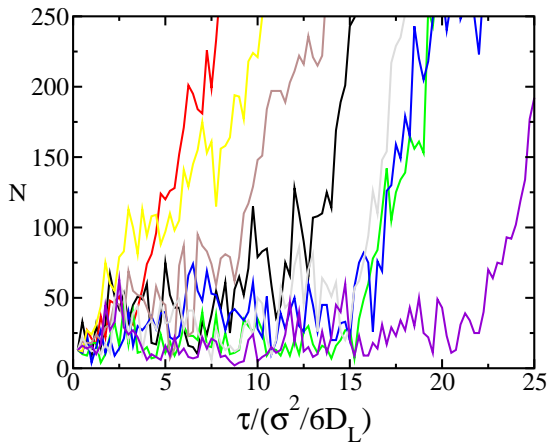}
	\caption{Number of particles belonging to the largest crystalline cluster versus time for different MD trajectories at $\phi$=0.52. $D_L$ is the long-time diffusion coefficient of the fluid 
spheres.}
\label{traj}
\end{center}
\end{figure}

\begin{figure}[htb]
\begin{center}
	\includegraphics[clip,width=0.4\textwidth]{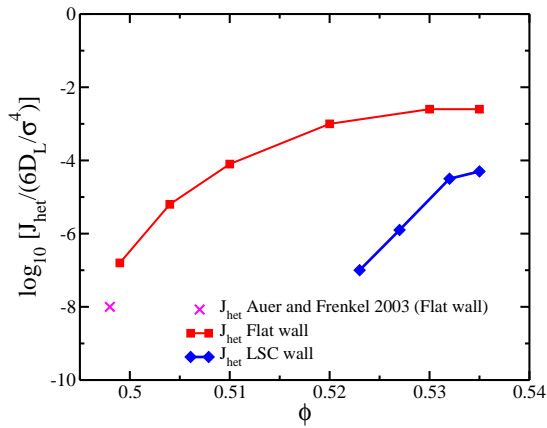}
	\caption{Red/blue: heterogeneous nucleation rate on a flat/LSC wall as a function of $\phi$. In pink we include the data
	obtained by Frenkel and Auer with Umbrella Sampling for a flat wall~\cite{auer2003line}.}
\label{jhetfig}
\end{center}
\end{figure}

In Fig. \ref{snap} we show a sequence 
of snapshots illustrating a heterogeneous crystallization event where a crystal cluster nucleates on the wall.  
In Fig. \ref{traj} we plot the time evolution of the number of particles in the largest crystal cluster (detected with a Steinhardt-like order parameter~\cite{PhysRevB.28.784,lechnerJCP2008} 
as implemented in Refs.~\cite{peterRoySoc2009,seedingvienes}). 
The figure shows different trajectories obtained at fixed $\phi$. 
Clearly, after an induction period, nucleation takes place and the number of particles with a crystalline environment (``crystalline particles'') rises sharply. 
To obtain the average time required to observe one heterogeneous nucleation 
event, $\left<t_{het}\right>$, we use some 8 trajectories 
for every pressure (each corresponding to a given choice of  $\phi$).
We then  obtain $J_{het}$ via Eq.~\ref{ref} (note that the area has to be multiplied by two to take into account 
that crystals can be formed in either side of the wall).
We repeat this procedure for several values of $\phi$ and obtain $J_{het}(\phi)$ as shown in Fig. \ref{jhetfig} in red. Our values are consistent
with those determined by Auer and Frenkel for low $\phi$ using Umbrella Sampling  (pink point in the figure)~\cite{auer2003line}.
Such consistency validates our simulation approach to obtain heterogeneous nucleation rates of hard spheres on hard flat walls. 
These results are not strongly sensitive to the interaction between the wall and the fluid particles: if such interaction
is changed from pseudo-hard sphere to Lennard Jones we obtain the same results within the uncertainty of our calculations. This is consistent 
with the fact that a small softness \cite{filionJCP2011} (or anisotropic shape \cite{ni:034501,zubieta2019nucleation}) do not change the nucleation rate very much.

\subsection{Homogeneous vs heterogeneous nucleation}

Once $J_{het}(\phi)$ is known, the average time required for a cluster to appear at the walls, $t_{het}$, can be obtained via 
Eq. \ref{ref}. We will compare this time with the one required for a critical cluster to appear in the bulk, $t_{hom}$:
\begin{equation}
t_{hom}=\frac{1}{V J_{hom}}.
	\label{thom}
\end{equation}
where $V$ is the volume and $J_{hom}$ is the homogeneous nucleation rate. We use a fit for $J_{hom}$ obtained in
simulations of seeded crystallization and inspired on Classical Nucleation Theory \cite{seedingvienes}. 
Such fit is consistent with independent calculations 
of the nucleation rate by means of other techniques like Umbrella Sampling, Forward-Flux Sampling and
brute-force simulations \cite{N_2004_428_00404,filion:244115,seedingvienes,richard2018crystallization}. 

\begin{figure}[htb]
\begin{center}
	\includegraphics[clip,width=0.4\textwidth]{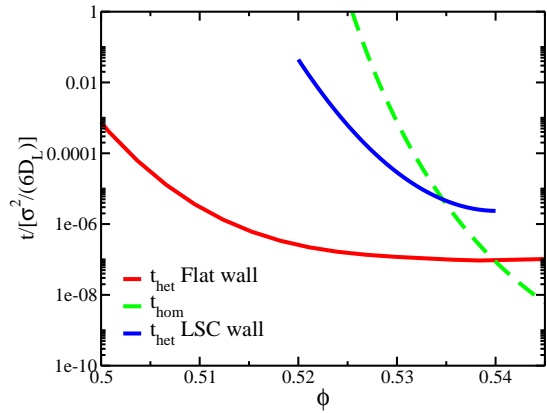}
	\caption{Estimates of the homogeneous (dashed green) and heterogeneous nucleation times as a function of $\phi$ in a 1x1x1 cm$^3$ 
	cell with flat hard walls (red) and LSC walls (blue) containing 200 nm radius particles.}
\label{times}
\end{center}
\end{figure}

From Eqs. \ref{ref} and \ref{thom}, it is evident that to estimate $t_{het}$ and $t_{hom}$ for a given sample-cell geometry, we 
need to specify  the volume and the surface area of the
sample cell. Here we consider a cubic cell of 1x1x1 cm$^3$ and particles of 200 nm radius, 
which are typical values for light-scattering experiments \cite{PRE_1997_55_3054,sinn,schatzel}. 
The resulting $t_{hom}(\phi)$ and $t_{het}(\phi)$ curves are shown 
in green and red respectively in Fig. \ref{times}. 
For a given $\phi$, the dominant mechanism is the one with the smallest nucleation time. 
Clearly, at low $\phi$'s heterogeneous nucleation is rapidly becoming dominant.
The curves cross
at $\phi \sim$ 0.54.
This is in fairly good agreement with experimental studies on the competition between homogeneous and 
heterogeneous crystallization in hard sphere-like colloids \cite{franke2011heterogeneous}.
Thus, our simulations suggest that, for flat walls 
and the system size under consideration (1x1x1 cm$^3$ and 200 nm radius particles) 
bulk crystal formation can only be observed for   $\phi \ge$ 0.54. 
The reason why there is a crossover between heterogeneous and homogeneous nucleation is that
$J_{hom}$  varies more sharply with $\phi$ than $J_{het}$. In Fig. \ref{jhetfig} it can be seen 
that $J_{het}$ only increases about 5 orders of magnitude in going from $\phi = 0.5$ to $\phi = 0.54$. 
By contrast, $J_{hom}$ increases by hundreds of orders of magnitude in the same density interval \cite{seedingvienes}.

\begin{figure}[htb]
\begin{center}
	\includegraphics[clip,width=0.4\textwidth]{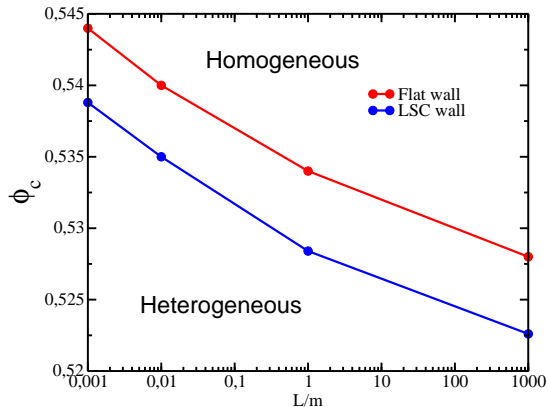}
	\caption{Crossover $\phi$ between heterogeneous and homogeneous nucleation as a function of the edge length of a cubic sample cell. 
	Red and blue correspond to flat and LSC walls respectively (see main text for details on the studied coating). 
	The curves can be interpreted as an indication of the minimum system size required to measure homogeneous nucleation 
	without the interference of crystallization at the walls for a given value of $\phi$. Particles
	with 200 nm radius are taken to perform this calculation. For any other particle size the $x$ axis should be accordingly rescaled by
	the factor $r_i/200$ nm, where $r_i$ is the radius of interest.}
\label{cross}
\end{center}
\end{figure}

We note that since $t_{hom}$ and $t_{het}$ depend on the volume and the area of the cell respectively,
the crossover $\phi$, $\phi_c$, is system size dependent. 
In Fig. \ref{cross} we plot $\phi_c$ versus the length of the edge of a cubic 
cell for particles of 200 nm radius. The red curve corresponds to a flat wall, which is the 
case we are discussing so far.   The $\phi_c$ curve can be interpreted as a prediction of the minimum 
system size required to measure homogeneous nucleation without the interference of crystallization at the walls for a given $\phi$. 
As expected from the discussion above concerning Fig. \ref{times}, a 1 cm$^3$ cell is needed for $\phi = 0.54$. 
For $\phi = 0.535$, already a 1 m$^3$  cell would be required. Thus, it is in practice not viable to  
measure homogeneous nucleation for volume fractions $\phi \le$ 0.535 using cells with flat walls.

\subsection{Coated walls}

Thus far we have considered heterogeneous nucleation on flat walls.
An obvious question 
is to what extent roughening the walls can suppress heterogeneous nucleation. 
A number of experimental studies on homogeneous nucleation have attempted to suppress heterogeneous 
nucleation by coating the walls of the sample cell  with sintered particles~\cite{ziese2013heterogeneous,S_2001_292_258}.
Such coating is meant to suppress the layering that is 
induced in the liquid by a flat wall.  Unfortunately, the works that explicitly mention the use 
of a coating~\cite{PRE_1997_55_3054,sinn,schatzel} 
focus on crystallization of charged rather than hard colloids.
 For instance, in the work by Ziese et al. the walls are coated with particles that 
had a diameter 3.4 times larger 
than those of the fluid particles~\cite{ziese2013heterogeneous}.
Inspired by this experimental work, we have repeated our heterogeneous nucleation study
with a coated wall built by 
randomly covering a plane with
non-overlapping hard spheres 3.4 times larger than the fluid ones (specifically, 142 spheres are located on a square which side is 14 times 
the diameter of a single fluid sphere).    
All spheres have their center on the same plane. We refer to this wall as ``large spheres coating'' (LSC).
The blue data in Fig. \ref{jhetfig} show that, for such a coating,  the heterogeneous nucleation rate goes
down by several orders  of magnitude. 
Consequently, the $t_{het}$ increases (in blue in Fig. \ref{times}) and the 
crossover density becomes smaller (in blue in Fig. \ref{cross}). However, 
the effect of the coating does not greatly alleviate the problem that
unrealistically large samples are required to eliminate the effect of heterogeneous
nucleation for volume fractions lower than $\phi=$0.525, where large discrepancies have
been found between experimental and simulation data on homogeneous nucleation 
rates \cite{Nature_2001_409_1020,filion:244115}. 

\begin{figure}[htb]
\begin{center}
	\includegraphics[clip,width=0.2\textwidth,angle=-90]{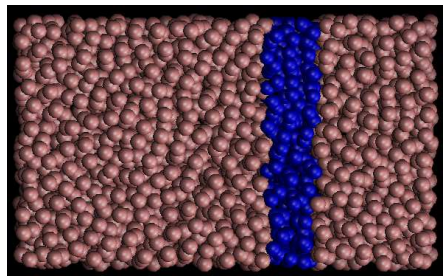}
	\caption{A fluid of hard spheres at $\phi=0.535$. Darker (blue) particles are frozen so
	that they form an amorphous solid wall with the structure of the fluid.}
\label{liquidwall}
\end{center}
\end{figure}

The question that arises now is whether there is an ideal coating that would 
suppress heterogeneous nucleation in experiments.  
An intuitive idea would be to use  an amorphous solid coating with the same structure as the liquid.
We refer to this kind of wall as ``frozen-liquid''. It seems plausible that the crystal--frozen-liquid 
interfacial free energy would be slightly higher than that of the 
crystal--liquid. The reason is that the frozen-liquid wall, unlike the liquid, 
is unable to accommodate the capillary fluctuations of the crystal--liquid interface
and the local structuring of a liquid close to a crystal interface.
Suppressing both effects by an external potential would require 
reversible work, and hence the crystal-frozen wall interfacial free 
energy is likely to be higher than that of the crystal-liquid interface. If 
that is the case, the crystal will not ``wet'' the frozen wall and 
therefore heterogeneous nucleation will not happen.
To test this idea 
we performed simulations  at $\phi = 0.535$ with a frozen-liquid wall
at the same density as the fluid, as shown in Fig. \ref{liquidwall}. 
The system in the figure is composed of 2548 fluid particles and the area 
of the simulation box side parallel to the solid wall 
is 11.5x11.5 $\sigma^2$. The average crystallization time in such system over 5 
trajectories is 12000$\pm 1000 \sigma^2/(6D_L)$. This time is virtually  
identical to that obtained with no wall and the same number of fluid particles: 11000$\pm 1000 \sigma^2/(6D_L)$. 
Moreover, we observe that crystal nuclei typically appear away from the frozen-liquid wall. In addition, when we double the system
size, leaving the area of the frozen-liquid wall fixed, the crystallization time is roughly halved. 
Therefore, it seems that under the conditions studied, the frozen-liquid wall is able to suppress heterogeneous nucleation completely.
We recall that on the LSC wall the crystallization 
time goes down from $\mathcal{O}(10^4)\sigma^2/(6D_L)$ to 75 $\pm 10 \sigma^2/(6D_L)$ because nucleation takes place heterogeneously.
With a flat wall crystal nucleation is  even faster (2$\pm 1  \sigma^2/(6D_L)$).
Both for the flat and the LSC walls, the observed crystallization time does not change when doubling the 
system volume while keeping the wall area constant.
In summary, in the presence of a ``frozen-liquid''  wall, we were able to reproduce homogeneous nucleation rates at $\phi=0.535$. 
Therefore, the use of a solid coating with the structure of the fluid could be a promising strategy to 
suppress heterogeneous nucleation in experiments. However, our study does not rule out the existence of other types of coatings also capable of 
effectively suppressing heterogeneous nucleation.

\begin{figure}[htb]
\begin{center}
	\includegraphics[clip,width=0.4\textwidth]{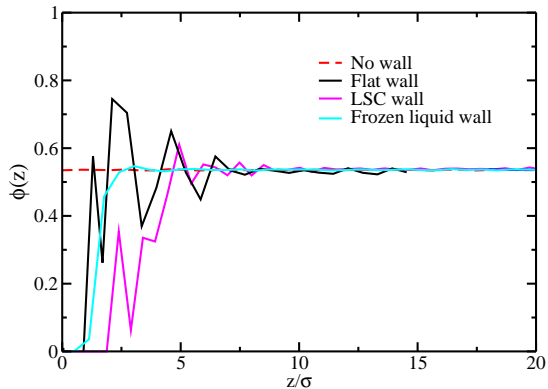}
	\caption{Fluid density profile along the direction perpendicular to the wall for flat, LSC and frozen-liquid walls for a fluid 
	at $\phi=0.535$. Density profiles start at zero from the middle of wall. 
	The fluid density profile under periodic boundary conditions (no wall) is also included for visual reference (dashed red).}
\label{denspro}
\end{center}
\end{figure}

\begin{figure}[htb]
\begin{center}
	\includegraphics[clip,width=0.4\textwidth]{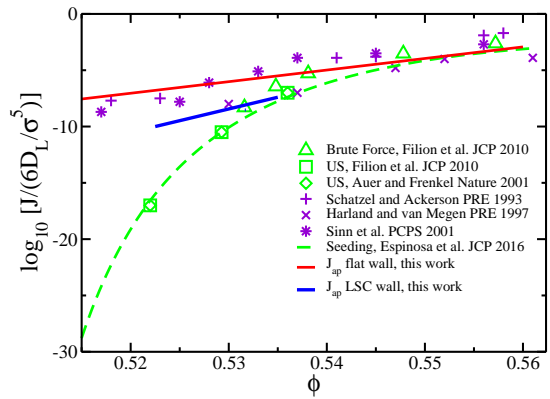}
	\caption{Crystal nucleation rate for the hard sphere system as a function of the volume fraction occupied by 
	the spheres, $\phi$. To compare experiments and simulations the rate is made dimensionless with the particle diameter, $\sigma$ and
	the long-time diffusion coefficient, $D_L$.
	Green and purple correspond to previous simulation and experimental studies respectively, 
	as indicated in the legend. Red (blue) line corresponds to our simulation estimate of the apparent homogeneous nucleation rate, which  
	takes both homogeneous and heterogeneous nucleation into consideration for a cell will flat (LSC) walls.}
\label{ratefig}
\end{center}
\end{figure}

Let us finish this section by comparing the structure of the fluid in all types of walls studied: flat, LSC and frozen-liquid.
In Fig. \ref{denspro} we show the fluid density profile along the direction perpendicular to the wall for $\phi=0.535$. For reference, we also include the
density profile of the fluid in a box with periodic boundary conditions (no walls). Clearly, the flat wall induces strong layering in the fluid, 
the LSC wall also induces an evident layering, although less pronounced, 
and the frozen-liquid wall does not induce any clear layering. Given that layering enhances crystal nucleation ability \cite{auer2003line}, 
the density profiles shown in Fig. \ref{denspro} help rationalise 
our results
that heterogeneous nucleation is readily induced by flat and LSC walls (by the former more efficiently than by the latter) and that
it is prevented by frozen-liquid walls.

\subsection{On the discrepancy between experiments and simulations}
Experimental measurements \cite{PRE_1997_55_3054,sinn,schatzel} and numerical calculations \cite{Nature_2001_409_1020,filion:244115} 
of $J_{hom}$  agree at high densities but differ by many orders of magnitude for lower values of $\phi$
\cite{Nature_2001_409_1020,filion:244115}. The situation is summarised in Fig. \ref{ratefig}, where simulation estimates and experimental measurements
of the rate are shown in green and purple respectively. 
The discrepancy between simulations and experiments is frustrating considering the simplicity of the system. 
However, on the basis of the heterogeneous nucleation rates that we have computed, we may reconcile experiment and simulation.
We define an apparent homogeneous nucleation rate, $J_{ap}$, that takes into account the simultaneous contribution
of homogeneous and heterogeneously nucleation:
\begin{equation}
J_{ap}=\frac{1}{t_{ap} V}
\end{equation}
where $V$ is the volume of the sample and $t_{ap}$ is the time required for the first critical cluster to appear, be it homogeneously
or heterogeneously. 
The inverse of $t_{ap}$ is the frequency of appearance of a critical cluster in the system. Such frequency will contain  
the contribution of two terms, one from homogeneously and another from heterogeneously nucleated crystallites:
\begin{equation}
\frac{1}{t_{ap}}=\frac{1}{t_{hom}}+\frac{1}{t_{het}}
	\label{qfreq}
\end{equation}
Using the numerical  data for $J_{hom}$ and $J_{het}$ we can estimate $t_{ap}$ and hence $J_{ap}$.
We use for our calculations particles of 200 nm radius and a 1x1x3 cm cell as in Ref. \cite{PRE_1997_55_3054}.
The resulting $J_{ap}$ is shown in Fig. \ref{ratefig} in red for flat walls and in blue for a random coating of larger spheres (size ratio 3.4). 
At large volume fractions ($\phi > 0.54$) the apparent rate curves coincide with the homogeneous nucleation 
curve from simulations (dashed green) and with experimental data (purple symbols). This is to be expected because  
the frequency of appearance of crystallites becomes increasingly dominated by the homogeneous contribution. 
At lower densities the heterogeneous contribution to the crystallization frequency in Eq. \ref{qfreq}
causes a departure of $J_{ap}$ curves from the $J_{hom}$ simulation curve.
The blue curve departs from the dashed green one at lower densities because the coating lowers
$J_{het}$ (see Fig. \ref{jhetfig}). In other words, randomly coating the walls with larger spheres widens the density range over which 
experimental measurements are not affected by heterogeneous nucleation, but only slightly. 
It is quite suggestive that the red $J_{ap}$ curve (flat walls) coincides with the experimental data from refs. \cite{PRE_1997_55_3054,sinn,schatzel}, 
where the cell walls were presumably flat (at least, these papers do not mention the use of any wall coating).
Of course, we cannot rule out that other factors, such as  sedimentation, incomplete shear melting, impurities, uncertainties in the 
determination of the volume fraction or hydrodynamic effects  also play a role in the experimental determination of the crystallization rate~ \cite{russo2013interplay,ketzetzi2018crystal,palberg2014crystallization}. However, we argue that a quantitative study of these more subtle effects can only be carried out once the experimental data have been properly corrected for wall-induced crystal nucleation. Our results suggest that applying such a correction
may go a long way towards reconciling the many orders of magnitude discrepancy between the homogeneous nucleation rate obtained in simulations and the apparent homogeneous nucleation rate observed in experiments.

\section{Summary and conclusions}
In summary,
we have used computer simulations to study the
competition between homogeneous and heterogeneous crystal nucleation 
in hard-sphere suspensions. 
To estimate the relative importance of homogeneous and heterogeneous nucleation
we have performed  systematic simulations of the rate of heterogeneous crystal nucleation for a range of different conditions, and combined the results with existing numerical data on homogeneous crystal nucleation.

We find a crossover from heterogeneous to homogeneous nucleation 
as the density increases. The crossover is due to 
the fact that the homogeneous nucleation rate increases more sharply with density 
than the heterogeneous rate. The crossover density depends on the system size and 
on the shape of the solid surface on which heterogeneous nucleation takes place. 
We find that, for the wall surfaces that we studied (flat, and randomly coated with spheres about 3 times 
larger than the suspended colloids)
avoiding heterogeneous nucleation for $\phi < \sim$ 0.535 is not feasible for reasonable sample sizes.
Therefore, we predict that the
studied walls would not avoid heterogeneous nucleation
in the density range where large discrepancies between
experimental measurements and simulation estimates of
the homogeneous nucleation rate have been found.
We propose an amorphous  coating based on the structure of the liquid that could effectively suppress heterogeneous 
nucleation. 

Finally, we show that an homogeneous nucleation rate
that takes into account both homogeneously and heterogeneously 
nucleated crystallites for cell sizes typically used in light 
scattering experiments,  coincides with the experimentally reported nucleation rate, provided that the walls are flat.
Thus, our work therefore suggests that heterogeneous nucleation at the cell walls may at least partly explain 
the reported discrepancy of many orders of magnitude between experimental measurements and simulation 
estimates of the homogeneous nucleation rate. 

\textbf{Acknowledgements}

This work was funded by grants FIS2013/43209-P and FIS2016/78117-P of the MEC.
J.  R. Espinosa acknowledges financial support
from the FPI grant BES-2014-067625 and the Roger Ekins Research Fellowship of Emmanuel College. 
E. S. and C. Valeriani thank F. Sciortino and P. N. Pusey for useful advices.   
The authors acknowledge the computer resources and technical assistance
provided by RES and the Centro de Supercomputacion y Visualizacion de Madrid (CeSViMa).
C.V. acknowledges funding from Grant FIS2016-78847-P of the MEC  and
from UCM/Santander (PR26/16-10B-2).

\bibliographystyle{ieeetr}

\end{document}